\makeatletter
\@ifundefined{@parse@version@dash}{%
\def\@parse@version#1{\@parse@version@0#1}
\def\@parse@version@#1/#2/#3#4#5\@nil{%
\@parse@version@dash#1-#2-#3#4\@nil}
\def\@parse@version@dash#1-#2-#3#4#5\@nil{%
  \if\relax#2\relax\else#1\fi#2#3#4 }
}{}
\makeatother

\documentclass[twocolumn, amsmath,amssymb, aps,pre]{revtex4-1}

\usepackage{graphicx}
\usepackage{dcolumn}
\usepackage{bm}
\usepackage{natbib}
\usepackage{hyperref}
\usepackage{multirow}
\usepackage{kotex}
\usepackage{color,xcolor}
\usepackage[super]{nth}
\usepackage{textcomp}
\usepackage{verbatim}
\hypersetup{breaklinks={true}}
\newcommand{\Lim}[1]{\raisebox{0.5ex}{\scalebox{0.8}{$\displaystyle \lim_{#1}\;$}}}

\begin{document}

\title{Growing Hypergraphs with Preferential Linking}

\author{Dahae Roh}
\author{K.-I. Goh}
\email{kgoh@korea.ac.kr}
\affiliation{Deaprtment of Physics, Korea University, Seoul 02841, Korea\\}

\date{\today}

\begin{abstract}
A family of models of growing hypergraphs with preferential rules of new linking is introduced and studied. The model hypergraphs evolve via the hyperedge-based growth as well as the node-based one, thus generalizing the preferential-attachment models of scale-free networks. We obtain the degree distribution and hyperedge size distribution for various combinations of node- and hyperedge-based growth modes. We find that the introduction of hyperedge-based growth can give rise to power-law degree distribution $P(k)\sim k^{-\gamma}$ even without node-wise preferential-attachments. The hyperedge size distribution $P(s)$ can take diverse functional forms, ranging from exponential to power-law to a nonstationary one, depending on the specific hyperedge-based growth rule. Numerical simulations support the mean-field theoretical analytical predictions. 

\end{abstract}

\maketitle

\section{\label{sec:Intro}Introduction}

Complex systems have been projected on to graphs, also called networks, and this projection has enriched our understanding on complex systems \cite{Newman2010,Barabasi2016NetSci}.  Graphs consist of nodes and edges, which respectively represent the elementary constituents (such as individuals in a social network) and their pairwise interactions. Despite the profound contributions, the network projection is fundamentally limited, because it neglects the potential group-wise interactions existing in many real-world complex systems, such as the group interactions in social networks \cite{Benson-2018-simplicial} and the multimeric complex-associated regulations in biological networks \cite{Klamt-PLoSCB-2009}. To address this issue theoretically, recently the concept of so-called higher-order interactions \cite{Battiston2020} is adopted to represent complex systems. Hypergraphs~\cite{Berge1989} and simplicial complexes \cite{Bianconi2021HON} are two major theoretical models for the higher-order interacting complex systems. Introduction of higher-order interactions has shown to qualitatively modify the collective phenomena such as the synchronization \cite{Sebastian2020}, the contagion \cite{Iacopini2019, Jhun2019,Guillaume2021,Jihye2023}, the core structure \cite{LeeJS2023arXiv}, and so on.

In this paper, we employ the hypergraphs as the theoretical platform for modeling the higher-order, groupwise interactions. Hypergraphs generalize the graphs by allowing the edges (called the hyperedges) to connect more than two nodes \cite{Berge1989}. A hypergraph ${\mathcal H}$ consists of the set of nodes ${\mathcal N}$ and the set of hyperedges ${\mathcal E}$, which respectively represents the individuals and their groupwise interactions. The number of nodes is denoted $N=|{\mathcal N}|$ and the number of hyperedges $H=|{\mathcal E}|$. The degree $k_i$ of node $i$ is the number of hyperedges the node belongs to; the size $s_h$ of hyperedge $h$ is the number of nodes which belong to the hyperedge. The size of largest hyperedges is called the rank of the hypergraph $r_{\mathcal H}=\mathrm{max}\{s_h\}$.

One of the hallmarks of the complexity of network systems is that real-world networks are often scale-free, meaning that the node degree distribution $P(k)$ follows an asymptotic power law $P(k)\sim k^{-\gamma}$. 
The documented examples of scale-free networks include the World Wide Web, the Internet, the citation networks, the collaboration networks, and the metabolic networks, to name a few \cite{Barabasi2016NetSci}.  Several foundational models \cite{Barabasi1999, Dorogovtsev2000PRL, Krapivsky2000PRL, Goh2001} have been proposed and studied to understand scale-free nature of empirical networks. 
A major family of such models is based on the so-called preferential attachment mechanism in the network growth process \cite{Barabasi1999}, meaning that the new node in a growing network tends to make a link to existing nodes not uniformly at random but with the probability proportional to the node's degree $k$.  

Given the extra degrees of freedom for hypergraphs, it is of interest to examine the scale-free architecture of higher-order interacting complex systems not only in terms of node degrees but also of hyperedge sizes. Analyses of existing datasets mostly of social interactions, both online and offline, reveal that by representing the groupwise interactions properly (that is, without splitting them into pairwise interactions), the resulting real-world hypergraphs are still often scale-free in the node degrees \cite{Patania2017,LeeYongsun2021}, whereas in hyperedge sizes one observes a broad spectrum of behaviors, from exponential-like short-tailed to power-law-like fat-tailed distributions \cite{Patania2017,Benson-2018-simplicial}. With this background, the main aim of this paper is to set up and study a family of basic models of growing hypergraphs which directly generalizes the growing network models with preferential linking \cite{Barabasi1999,Dorogovtsev2000PRL, Krapivsky2000PRL}. 

To generalize the growing network models into the growing hypergraph models, we make two key generalizations: i) in addition to the node-based growth, the hyperedge-based growth is included; ii) the concept of preferential attachment is generalized for the hyperedge-based growth. 
Specifically, we define the preferential choice of hyperedges in terms of hyperedge sizes. By studying a family of models with different combinations of growth rules, we show that by including the hyperedge-based growth, the growing hypergraphs can produce scale-free degree distribution $P(k)\sim k^{-\gamma}$, even without degree-preferential attachment, with the degree exponent $\gamma$ depending on the growth rules. The hyperedge size distribution $P(s)$, however, takes more diverse functional forms, ranging from exponential or power-law to nonstationary ones, depending on the specific hyperedge-based growth rule. 
We obtain the results analytically using the mean-field arguments as well as the master equation approach whenever possible, supported by extensive numerical simulations. 

Before proceeding, it would be appropriate to mention previous related works in the literature \cite{GenomeEvolution, NohJaeDong2005, Laurent2011, Wang2010EPJB, Owen2017, Kirill2021, LeeYongsun2021,Alexei2022arXiv, Barthelemy2022, Krapivsky2023JPA}.
These works can be classified into three major classes. The first and earliest \cite{GenomeEvolution, NohJaeDong2005, Laurent2011} are models which are based on the traditional notions of networks (that is, pairwise interactions only) but include group-based network evolutions. Although they use different language than hypergraphs, underlying concepts and mathematical tools are partly shared with the current work. Second and more recent \cite{Wang2010EPJB, Owen2017, Kirill2021, LeeYongsun2021} are the models of growing hypergraphs and simplicial complexes. Contrary to the current work, these models assume the evolution processes involving the groups of {\it fixed} size and thereby the bounded rank. 
Third and the most recent class of papers \cite{Alexei2022arXiv, Barthelemy2022, Krapivsky2023JPA} shares common theoretical rationale of hyperedge-based growth with our models, although specific implementation of theoretical ideas diverges. We summarize similarity and differences of these works and ours at the end of the paper. 

\section{Growing hypergraph models}
Following the growing network models, the growing hypergraph models we propose here also proceeds by a new node introduced in each `time' step, $N(t)=t+N_0$, where $N_0$ is the number of nodes in the initial hypergraph ${\mathcal H}_0$. We assume that each new node makes two hyperedges: i) a size-2 edge via node-based rule and ii) a hyperedge via hyperedge-based rule. With the former, node-based process alone, the models reduce to the growing network models. The hyperedge-based growth rule contains two key factors: i) the rule of choice and ii) the rule of formation. In the rule of choice, we generalize the concept of preferential attachment for the hyperedges: we define the {\it preferential choice of hyperedge} in terms of hyperedge sizes, that the probability $\Pi_h$ for a hyperedge $h$ to be chosen is proportional to its size $\Pi_h\propto s_h$. 
For the rule of formation, we study two limiting cases: i) the new node gets merged into the chosen existing hyperedge, in which case $H(t)=t+H_0$;  
ii) a new hyperedge $h'$ is created, which is the union of the chosen hyperedge and the new node. In this case $H(t)=2t+H_0$. We refer to the former rule as the $M$-model and the latter the $C$-model. In general,  both the limiting rules would occur concurrently, the combined effect of which can be straightforwardly addressed. These elementary growth processes are illustrated in Fig.~1. 

\begin{figure}
\begin{center}
\includegraphics[width=.99\linewidth]{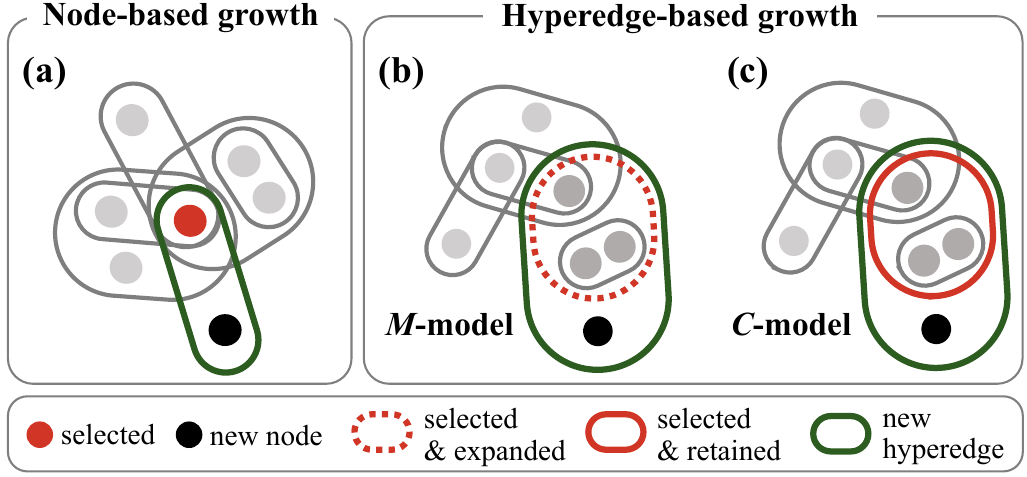}
\end{center}
\caption{Illustrations of the model rules. $\bold{(a)}$ Node-based growth. Hyperedge-based growth of $\bold{(b)}$ $M$-model and $\bold{(c)}$ $C$-model. Preferential selections are assumed.} 
\label{fig:Schematics}
\end{figure}

As the initial condition,  we take the initial hypergraph ${\mathcal H}_0$ to consist of two nodes connected by a hyperedge, that is $N_0=2$ and $H_0=1$. Algorithmically, we perform the following routines at each time step. 

\begin{itemize}
\item[\bf (i)] {\bf Node addition}: A new node is introduced. The number of nodes $N$ increases by one. 

\item[\bf (ii)] {\bf Node-based growth}: An existing node is chosen and a size-2 hyperedge is formed involving the new and the chosen node [Fig.~\ref{fig:Schematics}$\bold{(a)}$]. The number of hyperedges $H$ increases by one. 
The choice of existing node can be made either uniformly at random (denoted R) or preferentially (denoted P), defining two versions of the model. 

\item[\bf (iii)]{\bf Hyperedge-based growth}: Depending on the model (either model $M$ or $C$), do the following. 

\begin{itemize}
\item[\bf (iii)$^M$]{\bf Model $M$}: An existing hyperedge is chosen and the new node becomes the new member of the chosen hyperedge [Fig.~\ref{fig:Schematics}$\bold{(b)}$]. The number of hyperedges $H$ remains the same, and the size of the chosen hyperedge $s_h$ increases by one.

\item[\bf (iii)$^C$]{\bf $C$-model}: An existing hyperedge is chosen and a new hyperedge is formed, which is the union of the new node and the duplicate of the chosen hyperedge [Fig.~\ref{fig:Schematics}$\bold{(c)}$].
The number of hyperedges $H$ increases by one. 
\end{itemize}
The choice of existing hyperedge can be made either uniformly at random (denoted R) or preferentially (denoted P), defining two versions of the model each case. 

\end{itemize}
Node- and hyperedge-based growth processes are considered concurrent, {\it i.e.}, while conducting the rule (iii), the hyperedge generated in the rule (ii) is not encountered. Depending on the combination of rules of choice, four versions of each model are possible: $M_{\rm RR}$-, $M_{\rm RP}$-, $M_{\rm PR}$-, $M_{\rm PP}$-model (likewise for $C$-model), with the first subscript refers to the node choice in (ii) and the second the hyperedge choice in (iii). 

Before undertaking detailed theoretical analyses, let us begin with some reasoning about the expected properties of the models. From traditional network theory \cite{Barabasi1999, Dorogovtsev2000PRL, Krapivsky2000PRL} it is well-known that using the preferential attachment in node degree in rule (ii) alone is sufficient for the emergence of asymptotic power-law $P(k)$. Therefore, from the $P(k)$-perspective, the first question would be how, if ever, the hyperedge-based growth affects the power-law behavior of $P(k)$. Also of interest would be if the hyperedge-based growth alone can produce the power-law $P(k)$, to which we will give affirmative answer. 

Extending the insight from preferential-attachment models, one might expect the preferential choice of hyperedge would lead to scale-free $P(s)$ as well. This expectation turns out to be partly true, and we will show in which case the preferential hyperedge-based growth can give rise to power-law $P(s)$. 
Finally, empirical data analyses have shown that there are cases in which $P(k)$ and $P(s)$ exhibit qualitatively different forms, the former a power-law-like fat-tailed distribution yet the latter an exponential-like short-tailed one \cite{Patania2017}. In this regard, the possible origin of such duality is of interest, to which we give some insight.

\section{\label{sec:M-Models}Results: $M$-models}

In the $M$-models, each step only one new hyperedge with $s=2$ is added; we have $N(t)=t+N_0=t+2$ and $H(t)=t+H_0=t+1$. Each step the sum of node degrees in the system increases by 3 [2 from the new node and 1 from the existing node chosen in rule (ii)]; the sum of hyperedge sizes also increases by 3 [2 from the new hyperedge and 1 from the existing hyperedge chosen in rule (iii)$^M$]; we have the sum rule $Z_k(t)=\sum_jk_j(t)=Z_s(t)=\sum_gs_g(t)=3t+2$. 

\begin{figure*}[t]
\begin{center}
\includegraphics[width=0.99\linewidth]{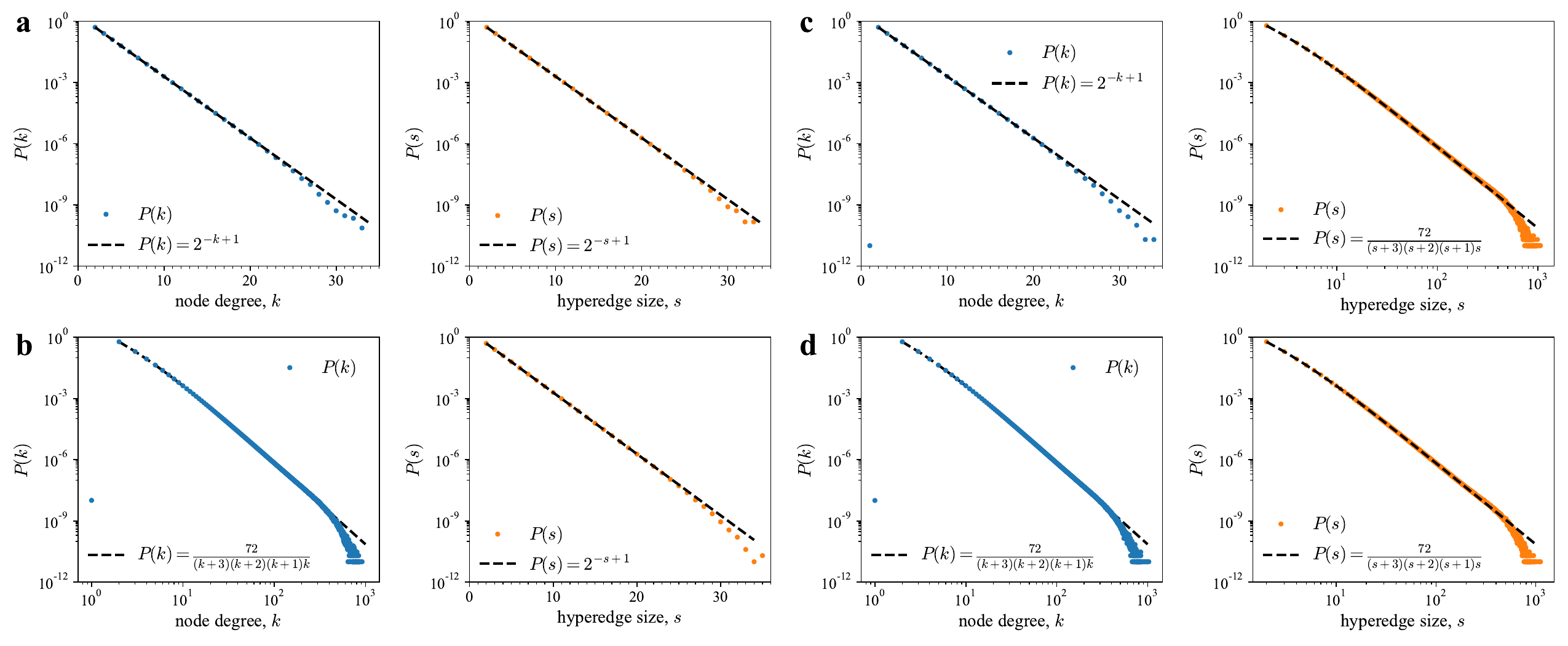}
\end{center}
\caption{The hyperdegree distribution $P(k)$ and hyperedge size distribution $P(s)$ for the $M$-model. Each plot corresponds to $\bold{(a)}$ $M_\text{RR}$-, $\bold{(b)}$ $M_\text{PR}$-, $\bold{(c)}$ $M_\text{RP}$- and $\bold{(d)}$ $M_\text{PP}$-models, respectively. Dashed lines denote analytic solutions and dots are for simulation results averaged over $10^5$ realizations with $N=10^6$.}
\label{fig:M_model}
\end{figure*}

In Fig.~\ref{fig:M_model}, the analytical solution for the resulting $P(k)$ and $P(s)$ are presented along with numerical simulation results, showing good agreement. As we shall see, in $M$-models the node-based and hyperedge-based growth effectively decouple and the each rule of choice fully determines the functional form of corresponding $P(k)$ or $P(s)$, respectively.

\subsection{Node degree distribution} 

Degree of existing nodes changes solely from rule (ii). Therefore, the degree distribution $P(k)$ of $M$-models is determined by the rule of node choice. For uniformly-random node-choice in rule (ii), $P(k)$ follows exponential distribution. The average number of nodes with degree $k$ at time $t$, $N(k,t)$, evolves as \cite{Dorogovtsev2000PRL, Krapivsky2000PRL}
\begin{equation}
N(k,t+1)=N(k,t)+\frac{N(k-1,t)}{t+2}-\frac{N(k,t)}{t+2}+\delta_{k,2}~. 
\label{eq:Pk_Mmodel_Random}
\end{equation}
Degree distribution at time $t$ is $P(k,t)=\frac{N(k,t)}{N(t)}=\frac{N(k,t)}{t+2}$. By assuming the asymptotic stationarity of degree distribution such that $P(k) = \lim_{t\to\infty}P(k,t)$ exists, one obtains the asymptotic stationary degree distribution as 
\begin{equation} 
P(k)=2^{-k+1} \left(k \geq 2\right)~~\mathrm{and}~~P(1)=0~.
\end{equation}

On the other hand, for the preferential node-choice, $P(k)$ follows an asymptotic power law. The rate equation for $k_i(t)$   \cite{Barabasi1999PhysicaA} is given by, using $Z_k(t)=\sum_j k_j(t)$, 
\begin{equation}
\frac{\partial k_i}{\partial t} = \frac{k_i}{Z_k(t)} = \frac{k_i}{3t+2},
\end{equation}
leading to $P(k) \sim k^{-\gamma}$ with $\gamma=4$. 
Evolution equation for $N(k,t)$ becomes 
\begin{equation}
N(k,t+1)=N(k,t)+\frac{(k-1)N(k-1,t)}{Z_k(t)}-\frac{kN(k,t)}{Z_k(t)}+\delta_{k,2}~, 
\end{equation}
with $Z_k(t)=3t+2$. 
Exact solution for asymptotic stationary $P(k)$ is obtained from master equation analysis \cite{Dorogovtsev2000PRL} as 
\begin{equation}
P(k) = \frac{72}{(k+3)(k+2)(k+1)k}\sim k^{-4}~.
\label{eq:Pk_Mmodel_Preferential}
\end{equation}
The master equation analysis is detailed in the Appendix. 
The degree exponent $\gamma=4$ differs from the degree exponent $\gamma=3$ for the original Barab\'asi-Albert model, which originates from the introduction of additional hyperedge-based growth.  

\subsection{Hyperedge size distribution} 

The size of existing hyperedges changes solely from rule (iii). Rule (ii) contributes only to the constant source term, describing the fixed size of the new hyperedge. Therefore, the size distribution $P(s)$ of $M$-models is determined by the rule of hyperedge choice. Furthermore, the governing equation for $P(s)$ evolution is essentially identical to that of $P(k)$ evolution. 

For uniformly-random hyperedge-choice, $P(s)$ follows exponential distribution. The average number of hyperedges with size $s$ at time $t$, $H(s,t)$ evolves as
\begin{equation}
H(s,t+1)=H(s,t)+\frac{H(s-1,t)}{t+1}-\frac{H(s,t)}{t+1}+\delta_{s,2}~.
\label{eq:Ps_Mmodel_Random}
\end{equation}
This equation differs from Eq.~(1) only slightly in the denominators, which is irrelevant in the long time (large $N$) asymptotic regime. 
Following the same steps as for the $P(k)$, one obtains 
\begin{equation}
P(s)=2^{-s+1} \left(s \geq 2\right)~.
\end{equation}
$P(s=1)=0$ by definition of the model.

For the preferential hyperedge-choice, $P(s)$ follows an asymptotic power law identical to Eq.~(5),
\begin{equation}
P(s) = \frac{72}{(s+3)(s+2)(s+1)s}\sim s^{-4}.
\label{eq:Ps_Mmodel_Preferential}
\end{equation}

The analytic results Eqs.~(2, 5, 7, 8) are compared with the numerical simulations results for $N=10^6$ in Fig.~2, all showing good agreement. 

\section{\label{sec:C-Models}Results: $C$-models}

In the $C$-models, each step two new hyperedges are added; one with $s=2$ from rule (ii) and the other from rule (iii)$^C$. The latter is created by `duplicating' an existing hyperedge and enlarging it with the new node: say, a publication by a new student joining an existing research team. It is noteworthy that the duplication-based growth underlies the gene-duplication models of protein network evolutions \cite{Sole2002ACS,Vazquez2003Complexus,GenomeEvolution,Goh2005JKPS}. 

In the $C$-models, we have $N(t)=t+N_0=t+2$ and $H(t)=2t+H_0=2t+1$. Each step the sum of node degrees in the system increases by $3+s_h$, where $s_h$ is the size of the chosen existing hyperedge, and so does the sum of hyperedge sizes; $Z_k(t)=\sum_jk_j(t)=Z_s(t)=\sum_gs_g(t)$ is no longer deterministic but itself a random variable in the $C$-models. 

\begin{figure*}[t]
\begin{center}
\includegraphics[width=.99\linewidth]{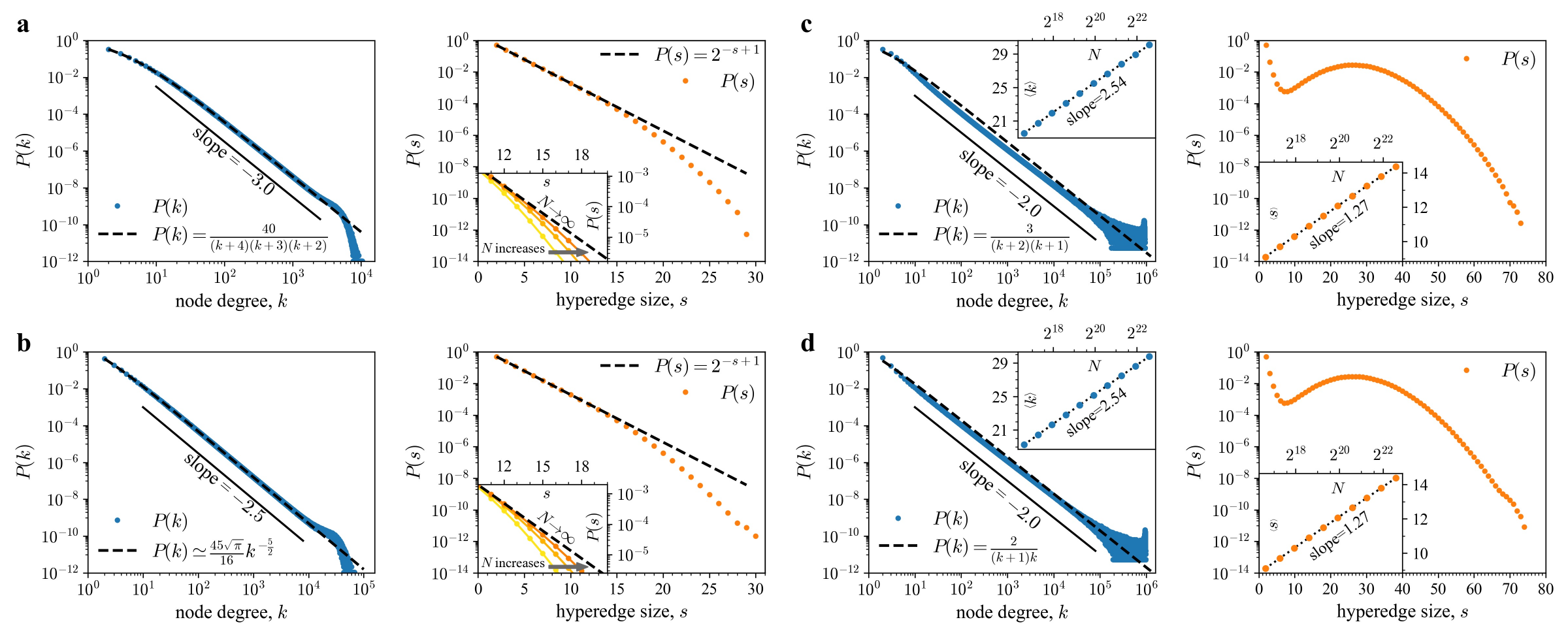}
\end{center}
\caption{Hyperdegree distributions $P(k)$ and hyperedge size distributions $P(s)$ for the $C$-model. Each plot corresponds to $\bold{(a)}$ $C_\text{RR}$-, $\bold{(b)}$ $C_\text{PR}$-, $\bold{(c)}$ $C_\text{RP}$- and $\bold{(d)}$ $C_\text{PP}$-model, respectively. The dashed lines are for analytic solutions and dots are for Monte-Carlo simulation results, averaged over $10^5$ realizations with $N=10^6$. {Insets in $\bold{(a)}$ and $\bold{(b)}$ display the system size dependence of and the approach to the asymptotic $P(s)$. Each line corresponds to $N=10^{4}$, $10^{5}$ and $10^{6}$, respectively. Insets in $\bold{(c)}$ and $\bold{(d)}$ are semilogarithmic plots of $\left<k\right>$ and $\left<s\right>$ with increasing $N$.} 
}
\label{fig:C_model}
\end{figure*}

In Fig.~\ref{fig:C_model}, the obtained $P(k)$ and $P(s)$ are presented. 
The node-based and hyperedge-based growth no longer decouple. The hyperedge-based growth plays dominant role in shaping the functional form of both $P(k)$ and $P(s)$: In particular, the degree distribution $P(k)$ asymptotically becomes a power law regardless of the rule of node choice, that is, even without the preferential attachment in node degrees. 

\subsection{Node degree distribution} 

\subsubsection{$C_\text{RR}$-model}
In this variation, both the existing node in rule (ii) and the existing hyperedge in rule (iii)$^C$ are chosen uniformly at random. 
Randomly choosing a hyperedge, however, gives rise to preferential selection from the perspective of nodes \cite{Dorogovtsev2001scalefree}.
Combining the two growth factors, one can write down the time evolution of node $i$'s degree as  
\begin{equation}
\frac{\partial k_i}{\partial t} = \frac{1}{N(t)} + \frac{k_i}{H(t)} = \frac{1}{t+2} + \frac{k_i}{2t+1}~,
\end{equation}
where the first and the second term on the right-hand side respectively accounts for the contribution of each growth process. The initial degree of every new node is fixed to be two. 
Accordingly, following \cite{Barabasi1999PhysicaA} one obtains $P(k) \sim k^{-\gamma}$ with $\gamma=3$ (Fig.~\ref{fig:C_model}$\bold{a}$). Asymptotic power-law $P(k)$ follows without the preferential attachment in node degrees. 

{To proceed with the master equation analysis for $P(k)$, it is noteworthy that for $C$-models the hyperedge-based growth induces significantly more involved contributions than for $M$-models, and the full master equation would become more involved. Instead, we employ a simplifying ansatz that the chosen hyperedge contains no more than one nodes with degree $k$, leading to an approximate master equation. This however does not compromise the accuracy of the solution significantly, as we will show. }

The approximate master equation for $P(k)$ of $C_{RR}$-model is written as follows:
\begin{align}
{N(k,t+1)\, } & {-\, N(k,t)=\frac{N(k-1,t)}{N(t)}-\frac{N(k,t)}{N(t)}}\nonumber\\
&{+\frac{(k-1)N(k-1,t)}{H(t)}-\frac{kN(k,t)}{H(t)}}+\delta_{k,2}~.
\end{align}
The left-hand side is the change in the average number with degree $k$ at time $t$. On the right-hand side, the first two terms are contributions from node selection, and the next two are from hyperedge selection. The last, source term is from the birth of a node. {The terms from hyperedge selection are approximate as discussed in the previous paragraph.} 
The stationary solution of Eq.~(10) is 
\begin{equation}
P(k) = \frac{40}{(k+4)(k+3)(k+2)}\sim k^{-3}~.
\label{eq:Pk_RR}
\end{equation}

\subsubsection{$C_\text{PR}$-model}
If a node is chosen preferentially, while a hyperedge at random, then the rate at which $i$-th node acquires degree becomes
\begin{equation}
\frac{\partial k_i}{\partial t} = \frac{k_i}{Z_k(t)} + \frac{k_i}{H(t)} = \frac{k_i}{\langle s\rangle (2t+1)} + \frac{k_i}{2t+1}~,
\end{equation}
where $\langle s\rangle = \sum_s sP(s)$ is the mean hyperedge size, which is time-dependent in general.
Below we shall show, however, that for the random hyperedge choice the ensemble average of $\langle s\rangle$ is a constant value $3$, independent of time. 
Mean-field-type approximation by replacing the denominator of the first term of the RHS by $3(2t+1)$ leads to the asymptotic behavior of $P(k) \sim k^{-\gamma}$ with $\gamma=5/2$ (Fig.~\ref{fig:C_model}$\bold{b}$).

The approximate master equation for $P(k)$ is given by
\begin{align}
\begin{split}
{N(k,t+1)\,}{-\,N}&{(k,t)=\frac{(k-1)N(k-1,t)}{Z_k(t)}-\frac{kN(k,t)}{Z_k(t)}}\\
&{+\frac{(k-1)N(k-1,t)}{H(t)}-\frac{kN(k,t)}{H(t)}+\delta_{k,2}}~.
\end{split}
\end{align}
The stationary solution can be obtained under the same mean-field approximation $Z_k(t)=3(2t+1)$ and reads 
\begin{equation}
P(k) = \frac{3}{2} \frac{\Gamma\left(\frac{1}{2} + 3\right)\Gamma\left(k\right)}{\Gamma\left(\frac{1}{2}+k+2\right)} \simeq \frac{45\sqrt{\pi} }{16} k^{-\frac{5}{2}}.
\label{eq:Pk_PR}
\end{equation}
Therefore, the inclusion of hyperedge-based growth rule preserves the scale-freeness of $P(k)$ but modifies the power-law degree exponent $\gamma$. 

\subsubsection{$C_\text{RP}$- and $C_\text{PP}$-model}
For $C_\text{RP}$-model, the rate at which $i$-th node acquires degree becomes 
\begin{equation}
\frac{\partial k_i}{\partial t} = \frac{1}{t+2} + \frac{\sum_{g \ni i} s_g}{\sum_h s_h} \approx \frac{1}{t+2} + \frac{k_i}{2t+1}\frac{\left<s^2\right>}{\left<s\right>^2},
\end{equation}
{where $\sum_{g \ni i}$ denotes a sum over every hyperedge $g$ where a node $i$ is involved and we replaced $\sum_{g\in i} s_g$ by its mean-field value $k_i\langle s^2\rangle/\langle s\rangle$.} 
Then we expect asymptotically
\begin{equation}
\gamma = \frac{2}{\Lim{t\to\infty}\frac{\left<s^2\right>}{\left<s\right>^2}}+1
\label{eq:gammaRP}
\end{equation}
with the convergence assumed. For $C_\text{PP}$-model,
\begin{equation}
\frac{\partial k_i}{\partial t} = \frac{k_i}{\sum_j k_j} + \frac{\sum_{g \ni i} s_g}{\sum_h s_h} \approx \frac{k_i}{2t+1}\frac{1}{\left<s\right>} + \frac{k_i}{2t+1}\frac{\left<s^2\right>}{\left<s\right>^2},
\end{equation}
\noindent{so we obtain similarly}
\begin{equation}
\gamma = \frac{2}{\Lim{t\to\infty}\left(\frac{\left<s^2\right>}{\left<s\right>^2}+\frac{1}{\left<s\right>}\right)}+1.
\label{eq:gammaPP}
\end{equation}
Exploiting the moment expansion of $P(s)$, the denominators of first terms in Eq.~(\ref{eq:gammaRP}) and (\ref{eq:gammaPP}) are obtained to converge to 2 in thermodynamic limit. Therefore, $P(k)$ is expected asymptotically to have a power-law tail as $P(k) \sim k^{-\gamma}$ with $\gamma=2$ in both $C_\text{RP}$- and $C_\text{PP}$-models, which is supported by Monte Carlo simulations (Figs.~3{\bf c},{\bf d}). 

The approximate master equation for $P(k)$ in $C_\text{RP}$-model is given by
\begin{align}
\begin{split}
{N(k,\,}&{t+1)-N(k,t)=\frac{N(k-1,t)}{N(t)}-\frac{N(k,t)}{N(t)}}\\
&{+\frac{\sum^{(k-1)}_{g} s_g}{Z_s(t)}N(k-1,t)-\frac{\sum^{(k)}_{g} s_g}{Z_s(t)}N(k,t)+\delta_{k,2}}~,
\end{split}
\end{align}
\noindent{{where the notation $\sum^{(k)}_{g}$ denotes the restricted sum over hyperedge $g$ containing a node with degree $k$. 
For $C_\text{PP}$-model, we have 
\begin{align}
\begin{split}
N(k, \, & t+1)-N(k,t)=\frac{(k-1)N(k-1,t)}{Z_k(t)}-\frac{kN(k,t)}{Z_k(t)}\\
&+\frac{\sum^{(k-1)}_{g} s_g}{Z_s(t)} N(k-1,t)-\frac{\sum^{(k)}_{g} s_g }{Z_s(t)}N(k,t)+\delta_{k,2}~.
\end{split}
\end{align}
To tackle the equations, we employ the mean-field approximation as $Z_s(t)=\langle s\rangle H(t)$ and $\sum_{g} s_g=k\langle s^2\rangle/\langle s\rangle$.
Under the approximation, the asymptotic stationary solution $P(k)$ for $C_\text{RP}$-model is obtained as 
\begin{gather}
P(k) = \frac{3}{(k+2)(k+1)}\sim k^{-2}~,\label{eq:Pk_RP}
 \intertext{and for $C_\text{PP}$}
 P(k) = \frac{2}{(k+1)k}\sim k^{-2}~. \label{eq:Pk_PP}
\end{gather}
Both the approximate solutions are in reasonable agreement with the numerical simulation results (Figs.~3c,d).

\subsection{Hyperedge size distribution} 

\subsubsection{$C_\text{RR}$- and $C_\text{PR}$-model}
If a hyperedge is chosen at random, the average number of hyperedges of size $s$ at time $t$, $H(s,t)$ would evolve as 
\begin{equation}
H(s,t+1)=H(s,t)+\frac{H(s-1,t)}{2t+1}+\delta_{s,2}.
\label{eq:P(e)_edgeR}
\end{equation} 
Asymptotic stationarity ansatz for the hyperedge size distribution $P(s) = \lim_{t\to\infty}P(s,t)=\lim_{t\to\infty}\frac{H(s,t)}{2t+1}$ finds the exponential hyperedge size distribution 
\begin{equation}
P(s)=2^{-s+1} ~~\left(s \geq 2\right)~,
\end{equation}
and $P(s=1)=0$. The mean hyperedge size becomes $\left<s\right>=3$. This asymptotic stationary $P(s)$ is of the same form as that of $M$-model, Eq.~(7),  despite the additional term in Eq.~(\ref{eq:Ps_Mmodel_Random}). At finite times (finite $N$), however, two models approach to the stationary exponential distribution differently. Specifically, the $C$-models converge to the asymptotics more slowly than $M$-models, as shown in  Figs.~\ref{fig:C_model}a,b. 

\begin{figure}[t]
\begin{center}
\includegraphics[width=0.99\linewidth]{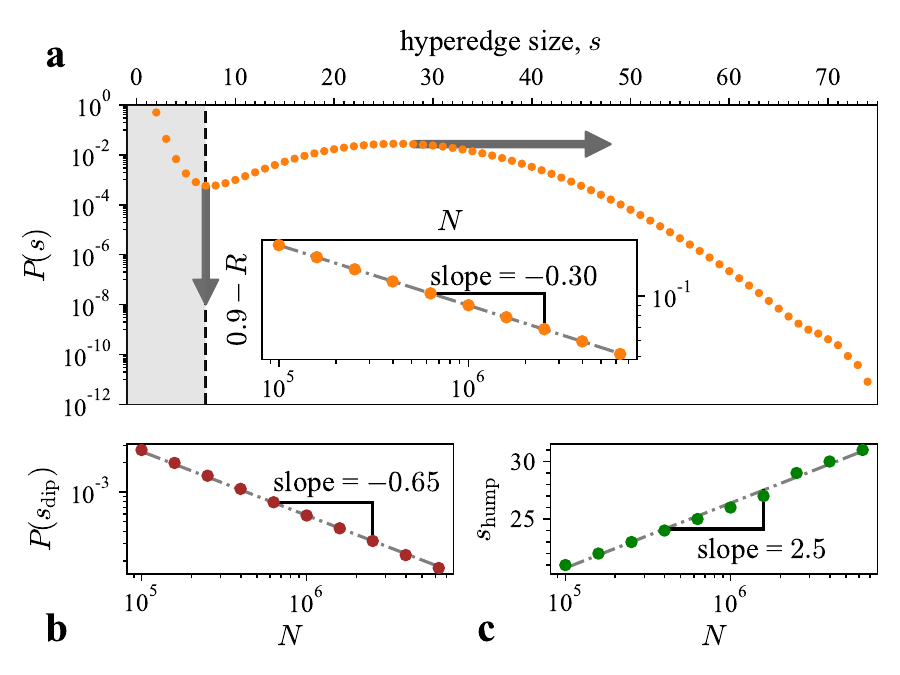}
\end{center}
\caption{
$\bold{(a)}$ Hyperedge size distributions $P(s)$ of $C_\text{PP}$-model with $N=10^6$ averaged over $10^5$ realizations. The vertical dashed line indicates the location of the dip, $s_{\rm dip}$, of $P(s)$.  Inset: Plotted is $R$ vs.\ $N$, where $R=\int_{s_{\rm dip}}^{\infty}P(s)ds\big/\int_2^{s_{\rm dip}}P(s)ds$, that is the ratio of the integrated statistical weight of $P(s)$ in $s>s_{\rm dip}$ (unshaded range) to that in $s<s_{\rm dip}$ (shaded range).  It is numerically obtained that $(0.9-R)\sim N^{-0.30}$. $\bold{(b)}$ The numerical value of $P(s_{\rm dip})$ at the dip of $P(s)$ in log-log scale in $N$. $\bold{(c)}$ Plot of $s_{\rm hump}$, the hyperedge size for the hump in $P(s)$ [marked by the horizontal arrow in (a)], in semilogarithmic scale in $N$. The dot-dashed lines indicate the linear-fit.}
\label{fig:C_edgeP}
\end{figure}

\begin{figure*}[t]
 \begin{center}
\includegraphics[width=0.9\linewidth]{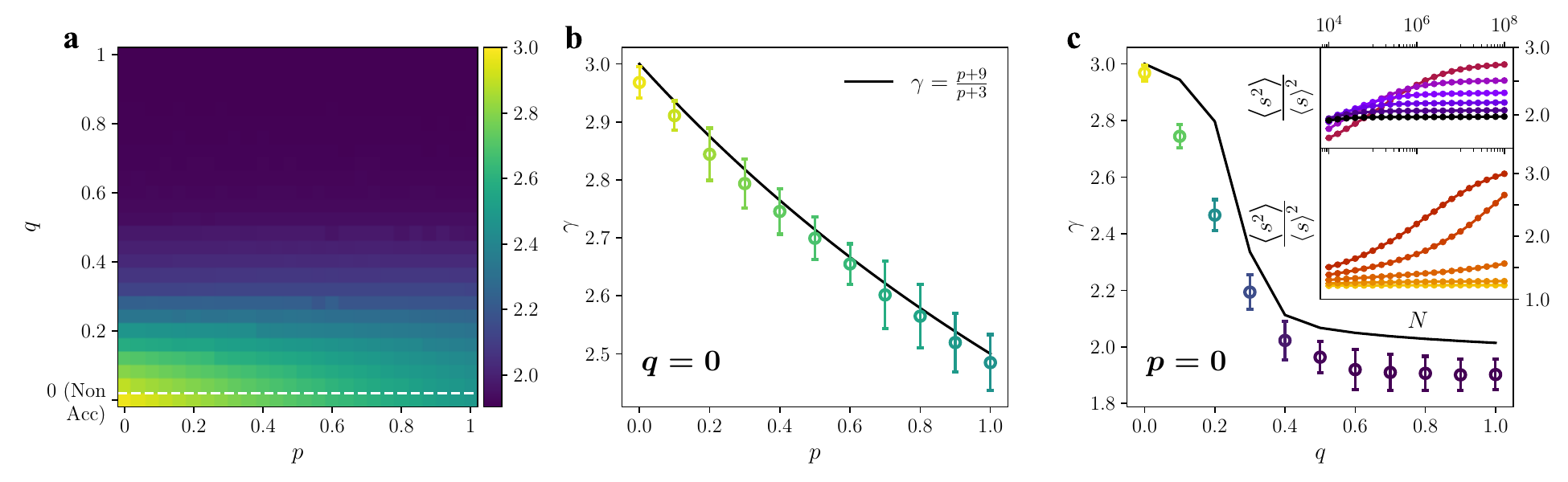}
\end{center}
\caption{$\bold{(a)}$ The degree exponent $\gamma$  for the probabilistic-mixture model. $p$ ($q$) is the probability to select a node (hyperedge) preferentially. Non-accelerating growth line ($q=0$) and accelerating growth region ($q>0$) are divided by a white dashed line. $C_\text{RR}$-, $C_\text{PR}$-, $C_\text{RP}$- and $C_\text{PP}$-model correspond to each corner. The value of $\gamma$ is estimated by linear-fitting the logbinned plot of $P(k)$, obtained from simulations with $N=10^6$ averaged over $10^4$ realizations. 
$\bold{(b)}$ Hyperedge-random (${q=0}$) and $\bold{(c)}$ node-random (${p=0}$) cases with standard deviations. Each case is compared with Eq.~(\ref{eq:gamma_edgeR}) and Eq.~(\ref{eq:gamma_nodeR}), respectively. In Eq.~(\ref{eq:gamma_nodeR}), ${\left<s^2\right>}/{\left<s\right>^2}$ at $N=10^8$ is used. Insets in $\bold{(c)}$ display  ${\left<s^2\right>}/{\left<s\right>^2}$ with increasing $N$ for different $q$'s. In the lower inset, from yellow to brown, $q=0$ to $0.4$ and in the upper inset, from brown to black, $q=0.5$ to $1.0$. Each point is averaged over $10^4$ realizations and the errorbars are smaller than the symbol size.} 
\label{fig:C_probabilistic}
\end{figure*}

\subsubsection{$C_\text{RP}$- and $C_\text{PP}$-model}

Under preferential choice of hyperedge, $H(s,t)$ evolves as 
\begin{equation}
H(s,t+1) = H(s,t) + \frac{\left(s-1\right) H(s-1,t)}{\sum_{s \geq 2} s H(s,t)}  + \delta_{s,2}~.
\end{equation}
{Analytic solution was not successful. The system experiences marginally accelerating growth \cite{Dorogovtsev2001accelerating} and $P(s)$ is strongly non-stationary. However, we still seek to understand some aspects of $P(s)$ through moments analysis and asymptotic behavior of it. }

By feeding the moments of $P(s)$, defined by $M_n = \sum_{s \geq 2} s^n P(s,t)$, into the master equation, we have 
\begin{equation}
\frac{\partial}{\partial t} \left[\left(2t+1\right)M_n\right] = 2^n + \frac{1}{M_1} \sum_{r=0}^{n} {n \choose r} M_{r+1}
\label{eq:Mn_edgeP}
\end{equation}
{for $n \geq 1$ and $M_0 =1$. In general, hierarchical equations---the equation for each $M_n$ containing a higher moment $M_{n+1}$---are insoluble \cite{krapivsky_redner_ben-naim_2010}.  Here, we compromise this problem by the linear growth of mean node degree and mean hyperedge size in $\ln{N}$, $\left<k\right> \sim \ln{N}$ and $\left<s\right> \approx \frac{1}{2} \langle k\rangle$, as depicted in the insets in Figs.~\ref{fig:C_model}(c,d). By substituting $M_0=1$ and $M_1 \sim \frac{1}{2} \ln{N}$, the leading term of each $n$-th moment of hyperedge size is obtained by induction as $M_n \sim \frac{1}{2} (\ln{N})^n$. Accordingly, the denominators of first terms in Eq.~(\ref{eq:gammaRP}) and (\ref{eq:gammaPP}) converge to two in thermodynamic limit neglecting the effect of node selecting rule.

Asymptotic behavior of hyperedge size distribution $P(s)$ in hyperedge-preferential model is studied numerically with $C_\text{PP}$-model (Fig.~\ref{fig:C_edgeP}). First, there remains the peak at $P(s=2)=\frac{1}{2}$ independent of $N$, because the node-selection rule keeps producing hyperedge of size 2. The rest of the distribution is shaped by the hyperedge-selection rule. Hyperedge size distribution $P(s)$ is divided into two areas by the vertical-dashed line in Fig.~\ref{fig:C_edgeP}a indicating the dip in $P(s)$. The probability weight of each side is comparable: the ratio of two areas divided by this vertical line approaches to $\approx$ 0.9 with increasing $N$ (inset of Fig.~\ref{fig:C_edgeP}a). With increasing $N$, $P(s_{\rm dip})$ decreases as a power law in ${N}$ (Fig.~\ref{fig:C_edgeP}b). The location of the hump, $s_{\rm hump}$ is shifting to the right logarithmically in ${N}$ (Fig.~\ref{fig:C_edgeP}c).
{The resulting $P(s)$ is strongly nonstationary, reminiscent of the nonstationary $P(k)$ observed in `densifying' networks studied in \cite{Densifying}. }

\subsection{Probabilistic-mixture model} 
We can interpolate the limiting cases of the $C$-model into probabilistic-mixture model (Fig.\ref{fig:C_probabilistic}), where a node (a hyperedge) is chosen preferentially with probability $p$ ($q$) or at random with probability $1-p$ ($1-q$).  Accordingly, the rate at which $i$-th node acquires degree becomes  

\begin{equation}
\frac{\partial k_i}{\partial t} =  \frac{1-p}{t+2} + \frac{p k_i}{\sum_j k_j} + \frac{\left(1-q\right)k_i}{2t+1} + \frac{q \sum_{g \ni i} s_g}{\sum_h s_h}.
\label{eq:ProbRateEq}
\end{equation}

\noindent{Monte-Carlo simulation results for varying $p$ and $q$ are presented in Fig.\ref{fig:C_probabilistic}$\bold{a}$. When $q\neq0$, system experiences accelerating growth \cite{Dorogovtsev2001accelerating}. Along the non-accelerating growth line ($q=0$), as described in random hyperedge-choice, hyperedge size distribution $P(s)$ follows stationary exponential function, independent of $N$ and $\left<k\right>$ and $\left<s\right>$ become also stationary as $\left<k\right> = 6$ and $\left<s\right> = 3$ in thermodynamic limit since $\left<k\right>=2\left<s\right>$. We further have}
\begin{equation}
\gamma = \frac{6}{p+3} + 1 = \frac{p+9}{p+3} \   \left(q=0\right)
\label{eq:gamma_edgeR}
\end{equation}

\noindent{and thus $\gamma$ varies from $3$ to $\frac{5}{2}$ as $p$ varies from $0$ to $1$ (Fig.~\ref{fig:C_probabilistic}b). One can further extend the range of $\gamma$ by introducing the initial attractiveness.}

For $q>0$, the problem becomes challenging because of the accelerating growth \cite{Dorogovtsev2001accelerating}  induced by preferential hyperedge selection. For $p=0$, we expect 
\begin{equation}
\gamma = \frac{1}{\frac{1-q}{2}+\frac{q\left<s^2\right>}{2\left<s\right>^2}} + 1,
\label{eq:gamma_nodeR}
\end{equation}
\noindent{from the same argument as in preferential hyperedge-choice. Monte-Carlo simulation and the conjecture based on Eq.(\ref{eq:gamma_nodeR}) are presented in Fig.~\ref{fig:C_probabilistic}c, showing qualitative agreement. The solution for general $p$ and $q$ is left as an open problem.}

\section{\label{sec:Conclusion}Summary and Discussion}

In this paper, a class of growing scale-free hypergraph models is studied generalizing the preferential attachment models \cite{Barabasi1999, Barabasi2000, Dorogovtsev2000PRL} of growing scale-free networks. The models presented consist of two elemental processes at each increment of time, which are node- and hyperedge-based growth processes. 
For group-based growth, two complementary rules called the $M$-model and $C$-model are studied. Each process could be either random or preferential. We have defined the hyperedge-preferential process to be preferential in the hyperedge size $s$, i.e. the number of nodes involved in the hyperedge. 
{The model variations investigated in this paper cover broad spectrum of functional forms of $P(k)$ and $P(s)$. Model variations with power-law $P(k)$ can be recategorized into three broad classes. The first is the cases with exponentially-bounded $P(s)$, including $M_{PR}$-, $C_{RR}$-, and $C_{PR}$-models; 
the second with also power-law $P(s)$, including $M_{PP}$-model; 
the third with nonstationary $P(s)$, including $C_{RP}$- and $C_{PP}$-model. }

{As mentioned beforehand, recent publications~\cite{Alexei2022arXiv, Barthelemy2022, Krapivsky2023JPA} share common theoretical rationale of hyperege-based growth with our models, although specific implementation of the theoretical ideas diverges. Here we briefly summarize the similarity and differences of these papers and our work. Vazquez proposed the notion of `natural' hypergraphs \cite{Alexei2022arXiv} which evolve through the duplication of uniformly chosen hyperedge and union of a new node with the duplicated one. 
Random recursive hypergraphs studied by Krapivsky \cite{Krapivsky2023JPA} evolve similarly.  
In fact, the $C$-model without the node-based growth process is equivalent to specific parameter instances of these models \cite{Alexei2022arXiv, Krapivsky2023JPA}. Barthelemy \cite{Barthelemy2022} proposed several ways to introduce preferential attachment in the hypergraph formation, including the one proportional to the hyperedge size, similarly to ours. The models studied in \cite{Barthelemy2022} however are not `growing' hypergraphs in the sense that number of nodes increases in time. Therefore, the statistical properties of the models are quite different from those of our models. 
}

The current models are meant to be minimal while keeping the essential growth ingredients of hypergraph evolutions. As such, several directions of generalizations and more detailed modelings are open. We mention some of them explicitly here:  
i) Structural correlations---{Intrinsic structural correlations \cite{Krapivsky2001PRE} in growing model can be another interests. It is anticipated that a node degree and a size of hyperedge where a node belongs are correlated via node age.}
ii) Partial duplications---We particularly focus on the limit where all the nodes in the pre-existing group (hyperedge) participate again in the new hyperedge. In many realistic situations, the duplication process might often be incomplete. One may model it by partial duplication processes akin to \cite{Sole2002ACS,Vazquez2003Complexus,Densifying}. A question of immediate interest is if and how the partial duplications can modify the nonstationarity of $P(s)$ for the hyperedge-preferential cases. 
iii) Finite lifetimes---A good example of node/hyperedge's finite lifetime would be neural interactions in brain activity. In the coauthorship network, retirement and subject-shift of an author may contribute to finite life time of nodes and hyperedges. Effects of such `aging'  \cite{Aging} can be of interest also in hypergraph evolutions. 
iv) Hyperedge overlaps---Hyperedge overlaps are higher-order structural correlations observed ubiquitously in real complex systems \cite{Palla2005, LeeGeon2021,Jungho2022arXiv} and can play a crucial role in higher-order dynamics \cite{Arruda2021arXiv}. It is thus of high interest which role the hyperedge overlaps may play in the hypergraph evolution. Introduction of more sophisticated evolution rules such as overlap-based growth process may be a promising topic to explore. v) Local events---Finally, more fine-tuned models of hypergraph evolution with various local events including creations and/or degradations of links between existing nodes can be formulated as in graphs~\cite{Barabasi2000}. Hyperedge-based local events may include the aggregation or dissociation of existing hyperedges \cite{Temporal_dynamics}.  }
Extension of the model in each of these directions is of ample interest and reserved for future works. 

\section*{Acknowledgement}
We thank L.~H\'ebert-Dufresne for insightful discussion. This work was supported in part by the National Research Foundation of Korea (NRF) grants funded by the Korea government (MSIT) (No. NRF-2020R1A2C2003669). D.R. is grateful for financial support from Hyundai Motor Chung Mong-Koo Foundation.

\appendix
\section{The master equation approach}

The master equation analysis due to \cite{Dorogovtsev2000PRL} has been utilized to obtain the exact solutions for power-law distribution and the exponent. In this appendix, we briefly outline how the method is applied in our models. 

Suppose at each time step a new node is introduced and makes $m$ directed connections by preferential attachments and $n_r$ directed connections by random attachments. After $t$ steps, the system consists of $t$ nodes and {$(m+n_r)t$} directed connections. {In the preferential attachment,} a node $i$ gains a new in-degree proportional to its {\it attractiveness}, $A_i = d_i+A$ where $d_i$ is the in-degree of node $i$ and $A$ is the {\it initial attractiveness}, a constant independent of $i$. 

One can write the master equation for the in-degree distribution $P(d,t)$ for this process, following \cite{Dorogovtsev2000PRL}, as 
\begin{widetext}
\begin{equation}
\left(t+1\right)P(d,t+1) -tP(d,t)  = n_r\Big[P(d-1,t)- P(d,t)\Big] + m\left[\frac{(d-1+A)P(d-1,t)  -(d+A)P(d,t)}{m+n_r+A} \right]+ \delta(d).
\label{eq:eq_master_P(d)}
\end{equation} 
The left-hand side is the change in the number of nodes with in-degree $d$ at time $t$. The first two terms on the right-hand side are contributions of random attachment, the next two terms are of preferential attachment and the last is a birth of a node. Using the Z-transform, $\Phi(z) = \sum_{d=0}^{\infty} P(d) z^d$, and the expansion of hypergeometric function~\cite{Dorogovtsev2000PRL}, the stationary solution for the distribution is obtained as 
\begin{equation}
P(d) = (1+a) \frac{\Gamma\left[(1+a)(1+n_r)+A\right]}{\Gamma\left[(1+a)n_r+A\right]} \frac{\Gamma\left[d+(1+a)n_r+A\right]}{\Gamma\left[d+(1+a)(1+n_r)+A+1\right]} ~,
\label{eq:P(d)_general}
\end{equation}
\end{widetext}
where $a=(A+n_r)/m$ and $\Gamma[\cdot]$ is the gamma function. The asymptotic behavior of $P(d)$ is of a power-law form, $P(d)\sim d^{-\gamma}$, with the power-law exponent given by
\begin{equation} \gamma=2+a~. \end{equation}

Different model variations are characterized by the parameters $m$, $n_r$, and $A$. 
First, throughout our models, $A=2$; note that the node degree $k$ relates to in-degree $d$ as $k=d+A=d+2$.
Now for $M$-models, we have both for $M_{PR}$- and $M_{PP}$-models, $m=1$ and $n_r=0$, leading to Eq.~(5);  
For $C_{RR}$-model, $m=\langle s\rangle=3$ and $n_r=1$, leading to Eq.~(11); For $C_{PR}$-model, $m=\langle s\rangle+1=4$ and $n_r=0$, leading to Eq.~(14); For $C_{RP}$-model, {$m=\langle s^2\rangle/\langle s\rangle$} diverges and $n_r=1$.  Therefore $a\to 0$, leading to Eq.~(21); 
For $C_{PP}$-model, {$m=\langle s^2\rangle/\langle s\rangle+1$} and $n_r=0$, leading to Eq.~(22). 
{The analysis for the hyperedge size distribution $P(s)$ can be applied in a similar way.}

\nocite{TitleOn}
\bibliography{HG_main}

\end{document}